\documentclass[letterpaper, 10 pt, conference]{ieeeconf}  

\IEEEoverridecommandlockouts                              

\usepackage{amsmath}
\usepackage{amsfonts}
\usepackage{siunitx}
\usepackage{graphicx}

\title{\LARGE \bf
Exploiting Weak Head Movements for Camera-based Respiration Detection
}

\author{Fabian Schrumpf$^{1}$, Christoph M{\"o}nch$^{1}$, Gerold Bausch$^{1}$ and Mirco Fuchs$^{1}$
\thanks{*This work was funded by the German Federal Ministry of Education and Research (BMBF) (FKZ 13FH032IX5)}
\thanks{$^{1}$Leipzig University of Applied Sciences (HTWK), Laboratory for Biosignal Processing}%
\thanks{\copyright 2019 IEEE. Personal use of this material is permitted. Permission from IEEE must be obtained for all other uses, in any current or future media, including reprinting/republishing this material for advertising or promotional purposes, creating new collective works, for resale or redistribution to servers or lists, or reuse of any copyrighted component of this work in other works.}
}

\begin{document}
\bstctlcite{IEEEexample:BSTcontrol}

\maketitle
\thispagestyle{empty}
\pagestyle{empty}

\begin{abstract}
In recent years, considerable progress has been made in the non-contact based detection of the respiration rate from video sequences. Common techniques either
directly assess the movement of the chest due to breathing or are based on analyzing subtle color changes that occur as a result of hemodynamic properties of
the skin tissue by means of remote photoplethysmography (rPPG). However, extracting hemodynamic parameters from rPPG is often difficult especially if the skin
is not visible to the camera. In contrast, extracting respiratory signals from chest movements turned out to be a robust method. However, the detectability of
chest regions cannot be guaranteed in any application scenario, for instance if the camera setting is optimized to provide close-up images of the head. In such
a case an alternative method for respiration detection is required.

It is reasonable to assume that the mechanical coupling between chest and head induces minor movements of the head which, like in rPPG, can be detected from
subtle color changes as well. Although the strength of these movements is expected to be much smaller in scale, sensing these intensity variations could provide
a reasonably suited respiration signal for subsequent respiratory rate analysis.

In order to investigate this coupling we conducted an experimental study with 12 subjects and applied motion- and rPPG-based methods to estimate the respiratory
frequency from both head regions and chest. Our results show that it is possible to derive signals correlated to chest movement from facial regions. The method
is a feasible alternative to rPPG-based respiratory rate estimations when rPPG-signals cannot be derived reliably and chest movement detection cannot be applied
as well.
\end{abstract}

\section{Introduction}
The respiratory frequency is an important indicator of a person’s health state. Deviations in breathing rate or breathing depth can be early warning signs of
physical disorders.

Traditional ways of measuring the respiratory rate are based on the attachment of a sensor system (e.g. face masks or chest straps) to the body \cite{grenvik1972impedance}.
These techniques are very accurate but also uncomfortable for the patients. Other, less invasive methods exploit variations in the electrocardiogram (ECG) or the
photoplethysmogram (PPG) that are correlated to respiration \cite{Charlton2016}. Although being slightly more comfortable, these sensor based measurements are not
suitable in situations where the attachment of sensors to the human body is undesired or impossible.

Recent advances in camera technology and computing power have led to the advent of remote rPPG. This technique uses standard RGB-cameras
to detect color variations in facial skin tissue. These color variations reflect hemodynamic properties and are mainly caused by the pulsatile component of the
pulse wave in the capillary vessels in the upper layers of the skin. Although the underlying effects causing these signals are not yet fully understood
\cite{Kamshilin2015}, it has been shown that color variations are most pronounced in the green channel of the RGB image \cite{Verkruysse2008}.

The rPPG has a waveform similar to the contact based PPG signal. Recent studies reported its use for respiratory rate estimation \cite{VanGastel2016}. Numerous
features of the rPPG signal can be used to estimate the respiratory signal: (I) baseline variations due to changes in intrathoracic pressure, (II) modulations of the pulse
amplitude due to changes in blood ejection volume and (III) modulation of the pulse rate caused by respiratory sinus arrhythmia. These features can be extracted
by means of various signal processing algorithms as. The rPPG based
respiration detection has its major limitation in the fact that skin regions have to be visible to the camera and must not be covered by, e.g., hair or even
textiles. Another issue with rPPG is the influence of physiological processes on the signal that are unrelated to respiration \cite{Allen2007}.

Respiration is also accompanied by characteristic movements of the chest. Corresponding respiration signals are usually determined by tracking characteristic
image structures in order to evaluate their trajectory over time \cite{Makkapati2017}. Sometimes the chest is not visible, e.g., if a person is lying
underneath a blanket in a hospital bed. Even in this case, the derivation of a respiration signal is still possible, for instance by analyzing the profile of the whole image
\cite{Bartula2013}.  However, such methods focus on large variations mainly evoked by chest
movement and, therefore, cannot be applied in situations when chest regions are not present in the recorded image. Such a situation occurs if the camera
setting is optimized to record close-up images of the head.

Since chest and head are physically coupled it is reasonable to assume that the movement of the chest also induces minor movements of the whole head. If that is
the case, respiration signals can be derived (1) from any region in the face even if it contains no visible skin and (2) from all color channels rather
than only from the green channel as in rPPG.

Here, we present a study to investigate whether the coupling between chest and head leads to reasonable suitable signals to estimate the respiratory frequency.
The participants were in a supine position and were asked to synchronize their respiration to a metronome. Videos covering the subjects face and upper chest
were recorded. The respiratory signal was derived by means of rPPG as well as from the raw camera signal (i.e., the raw pixel intensities) respectively by means
of the empirical mode decomposition (EMD) \cite{Huang1996}. The former method was only applied to facial regions, while the latter was applied to both chest and
facial (head) regions.

Respiratory frequencies showed good agreement with the metronome in both-pixel based as well as rPPG-based estimations. We conclude that the
pixel based estimation is suitable for both chest and facial areas. 

\section{Methods}
\subsection{Experimental setup}
\begin{figure}
\centering
\includegraphics[width=0.7\columnwidth]{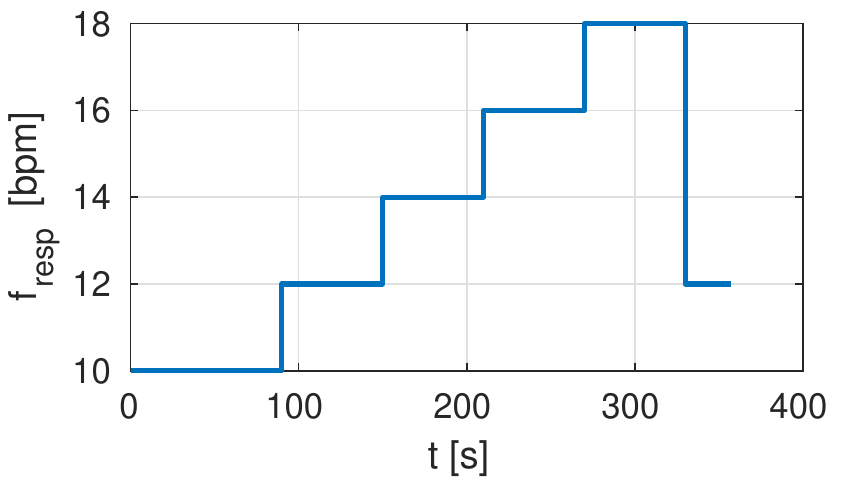}
\caption{Experimental protocol. The respiratory frequency was varied between
10 and 18 bpm.}
\label{fig:exp_protocol}
\end{figure}
We recruited subjects among healthy volunteers with a mean age of 28.75 years (SD 4.5 years). This study followed the tenets of the Declaration of Helsinki. Subjects were lying in a supine position and instructed to adjust their breathing frequency to a metronome. Respiratory frequency was varied in stages between
10 and 18 beats per minute (bpm) (see figure \ref{fig:exp_protocol}). Each subject completed between 5 and 12 breathing cycles per stage.
During the experiment, videos of the subject‘s face and upper chest were recorded using a standard RGB camera (IDS UI-3160CP, 120 fps, 1920 x 1080 pixel resolution).
The videos were stored in a raw file format for offline analysis using MATLAB R2017b. Subjects were instructed to move as little as possible to prevent motion
artifacts.
\subsection{Derivation of respiratory signals from RGB videos}
\begin{figure}
\centering
\includegraphics[width=0.7\columnwidth]{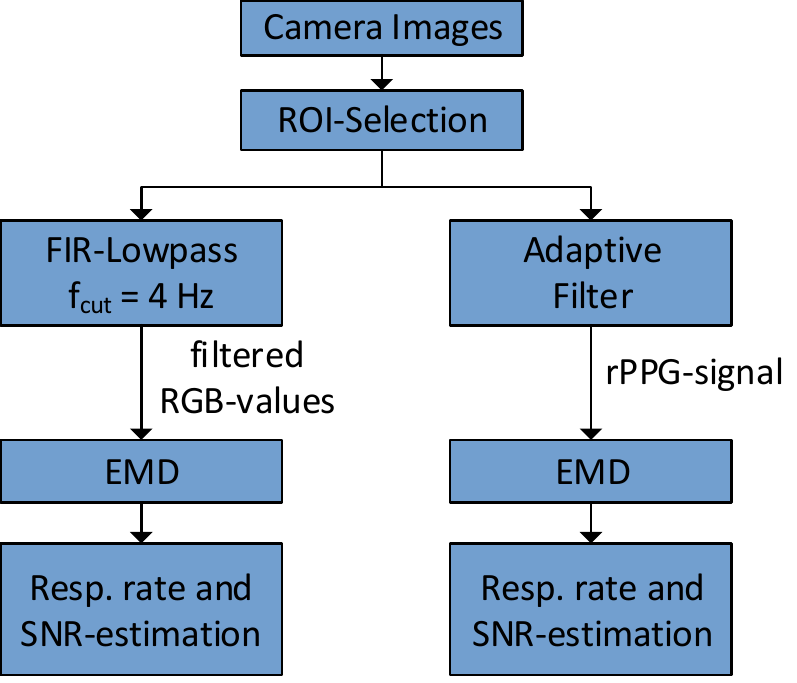}
\caption{Processing pipeline of the camera images recorded from the subjects' face and upper chest.}
\label{fig:proc_pipeline}
\end{figure}

In this study we aimed at deriving respiration signals using the following methods: (1) by means of parameters estimated from rPPG conducted in facial regions,
(2) by means of color variations of the raw pixel values in facial regions (i.e. movement signals), and (3) by means movements of the chest. The latter method
serves as a reference to demonstrate the feasibility of respiratory rate estimations based on raw color variations caused by body motion. The complete
processing pipeline is depicted in figure \ref{fig:proc_pipeline}. The signals were derived as follows.

First, a rectangular area of the subjects face and upper chest was selected as the main region of interest (ROI). This area was further divided into
square-shaped sub-ROIs with an edge length of 10 pixels. Sub-ROIs not belonging to the subject's face (e.g. background, hair) were excluded from further analysis by
manual background segmentation.

In order to extract the motion-based respiratory signals from both face and chest regions, the pixel values of each color channel were averaged in each valid
sub-ROI. These averaged values were subsequently low-pass filtered (FIR filter, \(f_{cut} = \SI{4}{\hertz}\)) to suppress high frequency noise. Sub-ROIs
located in the face provide information on facial movements due to respiration. Likewise, sub-ROIs in the chest area provide the chest signals. Note that we did not employ a
tracking based method to sense movements in the chest area since we do not expect such methods to work in facial areas. However, a direct comparison of the
movement derivation in both face and chest was desired.

The rPPG signal in each sub-ROI was calculated using a least mean square adaptive filter with the averaged red and green color channels as input signals to
suppress global intensity changes resulting from, e.g., varying lighting conditions \cite{Li2014}.
\subsection{Estimation of the respiratory rate}
To estimate the respiratory rate the time courses of the raw RGB values as well as the rPPG signal were divided into windows with a length of \SI{30}{\second}
and a step size of \SI{1}{\second}. In each window the EMD of every signal in each sub-ROI was computed \cite{Huang1996}. The EMD decomposes a signal into
several intrinsic mode functions (IMF) that form an almost orthogonal basis. Similar to the Fast Fourier Transform, the IMF represent the oscillatory modes that are embedded in the
signal. Given the fact that the respiratory information is modulated onto the color- and rPPG-signals, one of the IMF can be treated as a surrogate for the
respiratory signal. To identify that specific IMF, the power-spectral-density (PSD) of every IMF was computed and the frequency component with the highest
amplitude was selected as the representative frequency of the IMF. Subsequently, the IMF with a representative frequency within a plausible range was chosen as
the respiratory signal \cite{Madhav2011a}. The range of plausible respiratory frequencies was set from \SI{0.1}{\hertz} to \SI{0.4}{\hertz} \cite{Lindberg1992}.
The estimated respiratory frequency \(f_{k}\) (\(k = 1 \ldots N_{ROI}\)) of each sub-ROI was computed using the auto-correlation function of the respective respiratory signal \cite{Schrumpf2016a}.

Furthermore, the signal-to-noise ratio (SNR) was computed from the PSD of every respiratory signal using equation \ref{eq:SNR}. 
\begin{equation}
SNR_k = \frac{\sum_{f=\SI{0.1}{\hertz}}^{\SI{0.4}{\hertz}}P_k(f)}{\sum_{f=0}^{\SI{4}{\hertz}}P_k(f)}
\label{eq:SNR}
\end{equation}
The respiratory rate of each time window was
finally estimated as the weighted median of all sub-ROIs. To that end, the frequencies \(f_k\) were ordered and associated with a weight \(w_k\). The weights
were calculated from the SNR values using equation \ref{eq:weights}.
\begin{equation}
w_k=\frac{SNR_k}{\sum_{k=1}^{N_{ROI}} SNR_k}
\label{eq:weights}
\end{equation}
The weighted median of the respiratory rates was the value \(f_k\) that satisfied the condition in equation \ref{eq:w_median}.
\begin{equation}
\sum_{k=1}^{i-1}w_k \leq 0.5 \lor \sum_{k=i+1}^{N_{ROI}}w_k \leq 0.5
\label{eq:w_median}
\end{equation}
The respiratory rate \(f_k\) according to equation \ref{eq:w_median} was then compared to the respiratory frequency \(f_{resp}\) given by the metronome in order
to assess the accuracy of the estimation.

\section{Results}
\begin{figure}[t]
\centering
\includegraphics[width=0.75\columnwidth]{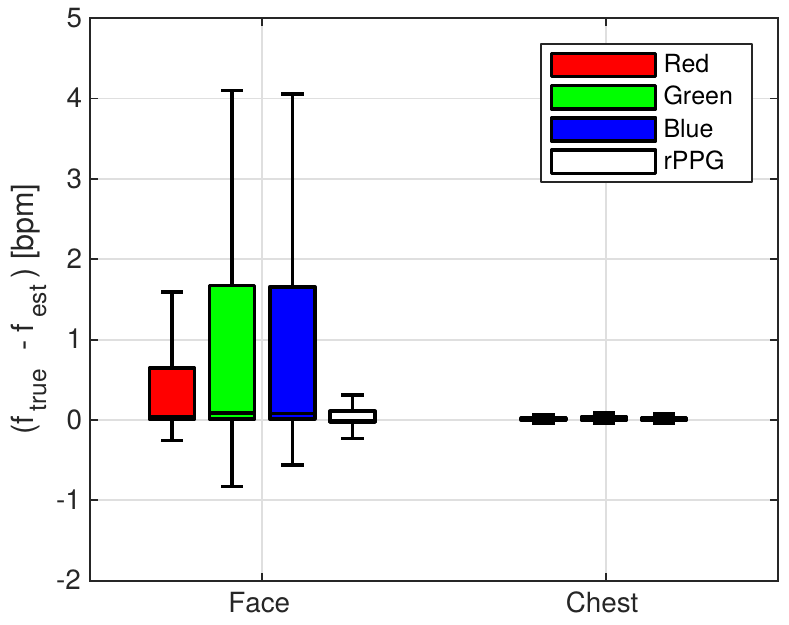}
\caption{Deviation of the estimated respiratory rate from the true respiratory rate based on raw pixel values and rPPG-signals on the face. For the chest
region only the RGB channels are displayed.}
\label{fig:err_resp}
\end{figure}
Figure \ref{fig:err_resp} shows the deviation of the estimated respiratory rate based on raw pixel values and the rPPG signal from the ground truth. The
extraction of the respiratory rate from the chest based on raw pixel values, which served as a reference, yielded almost perfect results

The best results in facial regions were obtained for rPPG. The movement based signals from green and
blue color channels have a similar median error as the rPPG based results but also a significantly greater variance. The red channels showed the lowest error variance of all color channels. However, error values in
the face were generally greater than on the chest and ranged between -0.25 bpm and 0.5 bpm.

\begin{figure}[t]
\centering
\includegraphics[width=0.75\columnwidth]{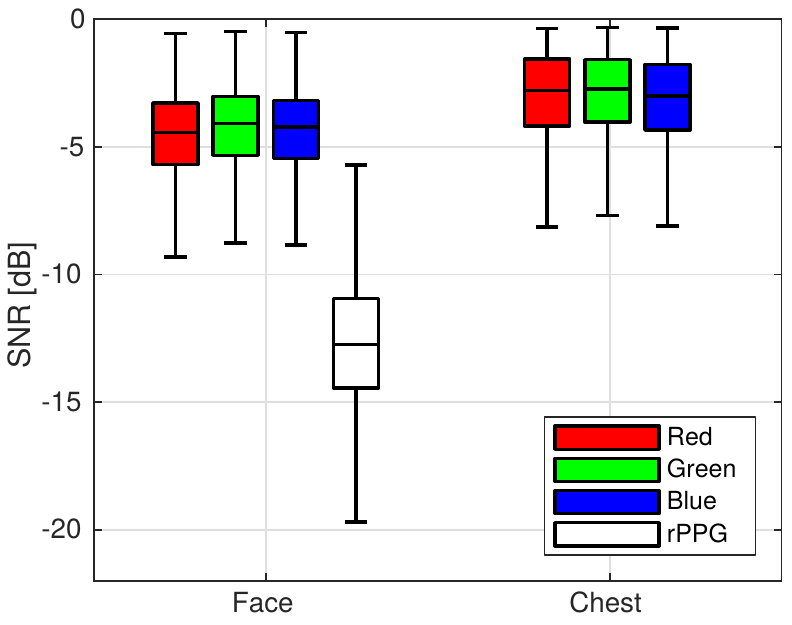}
\caption{SNR of the respiratory signal derived from the RGB and rPPG signals. SNR values for rPPG from the chest region have
been omitted.}
\label{fig:resp_SNR}
\end{figure}

To further investigate the sources of the respiratory signal we computed the SNR of the respiratory signal embedded in the raw pixel values and the rPPG signal.
Figure \ref{fig:resp_SNR} shows those SNR values for chest and facial regions separately. It can be seen that the color signals on the chest exhibited the highest SNR
with a median of \SI{-2.5}{\decibel}. This is due to the fact that movements associated with breathing are most pronounced on the chest.

In the facial region, the raw color signals showed an SNR ranging from \SI{-11}{\decibel} to \SI{-0.5}{\decibel} with the red color channel showing the highest median
SNR.
These values were lower than the SNR-values obtained on the chest. Extracting the rPPG-Signal from these channels lead to a drop in SNR. The rPPG-SNR ranged
between \SI{-17}{\decibel} and \SI{-0.5}{\decibel} with a median at \SI{-7}{\decibel}.

\begin{figure*}[t]
\centering
\includegraphics[width=0.8\textwidth]{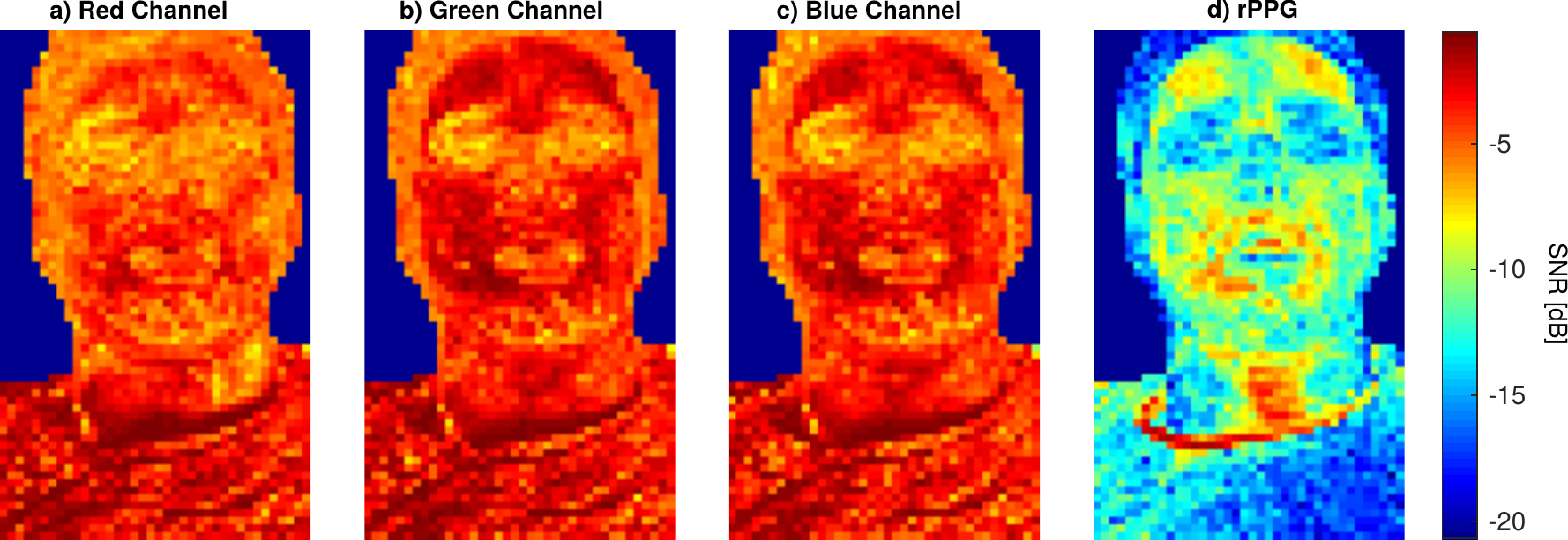}
\caption{Color mapping of the SNR-values of every sub-ROI. Background areas were not considered for
SNR estimation.}
\label{fig:subject_SNR}
\end{figure*}

The difference of the SNR values when comparing the four signal sources is further emphasized by figure \ref{fig:subject_SNR}. It is evident that the respiratory signal is most
pronounced in the raw pixel values (figure \ref{fig:subject_SNR} a-c). The SNR is highest on the chest where changes in pixel brightness caused by breathing movements can
be detected easily. The computation of the rPPG-Signal using an adaptive filter causes a drop in SNR (figure \ref{fig:subject_SNR} d). However, there are still regions with high
SNR-values in the face. These regions are primarily located on the cheeks and on the forehead. These are also the regions with the highest blood perfusion of
the upper skin layers.
\section{Discussion}
In the presented study we compared the accuracy of video image based respiration rates estimated from (1) rPPG signals derived from facial regions, (2) movement
signals derived in head regions (i.e., the face as well) and (3) movement signals derived from chest regions. The movement signals were directly calculated from
variations of the pixel intensities in the raw RGB color channels. In contrast, rPPG was calculated as commonly done only from the green channel. Moreover, global
intensity variations as they occur due to, e.g., tiny movements were compensated by means of an adaptive filter. Of course, this is only of minor importance for
the respiration rate estimation that exploits the respiratory sinus arrhythmia. However, it restricts the acquisition of respiration signals to skin regions. In contrast, the movement signals can effectively recorded in any head region and would therefore also work if skin cannot be detected at all. For
example, this might be the case when a subject lies prone.

It should be mentioned that rPPG is sometimes also referred as averaging pixel values (usually the green channel) within a ROI without further processing like
applying an adaptive filter \cite{Trumpp2017}. Our approach turns out to be fairly similar to such methods, although we exploit all color channels. However,
we are focusing on deriving movements that can also be recorded in non-skin regions and the calculated signals should therefore not be confused with rPPG
signals.

We conducted a study to analyze whether respiration related movements of the chest can also be measured from head regions. We expected this to be possible due
to the mechanical coupling between chest and head. A comparison of the estimated respiratory rates with the ground truth provided information on the suitability
of motion and rPPG-based approaches.

It has been demonstrated that the estimation of the respiratory rate from the chest based on RGB values yielded almost perfect results. This is due to the fact
that movements associated with breathing are strongest on the chest and are often visible to the naked eye.

On the face, rPPG yielded the best results. As stated earlier the pulse wave derived from the upper layers of the facial skin is modulated by respiration in
several ways. This modulation can be easily extracted by means of the EMD. The accuracy of the respiratory frequency based on RGB-signals resulted in larger
errors. Raw RGB values are corrupted by motion artifacts and disturbances caused by changes in illumination. However, movement information is still embedded in
the signal. Comparing the results of the three color channels revealed that the red channel performed better than the other two channels. The
utilization of sensor fusion methods may be a way to improve the robustness of the estimates derived from the RGB channels.

The respiratory frequency is an important tool for the assessment of a person’s health state. Therefore, the application of effective algorithms is of utmost
importance. The present study has shown that the camera-based derivation of the respiratory rate from a patients face based on motion and the EMD is feasible.
The exploitation of movement information enables the utilization of body regions that contain no visible skin. Furthermore, the extraction of additional signal
features (e.g. by means of peak detection, baseline estimation) is not necessary. By using the EMD the oscillatory mode related to respiration is calculated
directly from the signal. Further research will be directed towards the influence of the body position on the respiratory rate extraction.

\bibliographystyle{IEEEtran}
\bibliography{IEEEabrv,literature}
\end{document}